\documentclass[prl,showpacs,amsmath,twocolumn]{revtex4}  
\newcommand{\beq}{\begin{equation}}
\newcommand{\enq}{\end{equation}}

\usepackage{graphicx}

\begin{document}
\title{Collapse of Bose-Einstein condensate with dipole-dipole interactions}

\author{Pavel M. Lushnikov$^{1,2}$
}

\affiliation{$^1$ Theoretical Division, Los Alamos National Laboratory,
  MS-B284, Los Alamos, New Mexico, 87545
  \\
  $^2$ Landau Institute for Theoretical Physics, Kosygin St. 2,
  Moscow, 117334, Russia
  }
\email{lushnikov@lanl.gov}


\begin{abstract}
A dynamics of Bose-Einstein condensate of a gas of bosonic particles with long-range dipole-dipole
interactions in a harmonic trap  is studied. Sufficient analytical criteria are found both for
catastrophic collapse of Bose-Einstein condensate and for long-time condensate existence.
Analytical criteria are compared with variational analysis.
\end{abstract}


 \maketitle


Bose-Einstein condensation  of dilute trapped atomic gases
\cite{Anderson1995,Bradley1997} essentially depends on the
interpaticle interactions. In most experiments so far the
dominated interactions were short-range van der Waals forces which
are characterized by the $s-$wave scattering length $a$.
Spatially homogeneous condensates with positive scattering length
(repuilsive interaction) are stable while condensates with
negative scattering length (attractive interaction) are always
unstable to local collapses \cite{PitaevskiiRevModPhys1999}
because the quantum pressure is absent in homogeneous
condensates. The presence of trapping field allows to achieve a
metastable Bose-Einstein condensate (BEC) \cite{Bradley1997} for
$a<0$ if the number of particles is small enough to ensure
existence of local minima of energy functional
\cite{PitaevskiiRevModPhys1999}.

Recent progress in creating of ultra-cold molecular clouds
\cite{JDWeinstein1998,Wynar2000} opens a new prospective for
achieving BEC in a dilute gas of polar molecules and stimulates
growing interest in study of BEC with dipole-dipole interactions
\cite{Yi2000,Goral2000,santosPRL2000,Yi2001,Yi2002,Martikainen2001,Pu2001}.
Dipole-dipole forces are long-range and essentially anisotropic.
Net contribution of dipole-dipole interactions can be either
repulsive (positive dipole-dipole interaction energy) or
attractive (negative dipole-dipole interaction energy) depending
on the form of condensate cloud, its orientation relative to
dipole polarization axes and trap geometry. Respectively
stability and collapse of BEC strongly depends on clouds
anisotropy which opens a whole bunch of new phenomena to be
observed and makes task of achieving and control of BEC
especially challenging.

Dipole-dipole interactions can dominate provided polar molecules are oriented by strong enough
electric field. Similar effects can be achieved for ground-state atoms with electric dipole moments
induced by a strong electric field \cite{Yi2000,Yi2001}. Another possible physical realization is
atoms with laser induced electric dipole moments \cite{santosPRL2000}. Dipole-dipole interactions
can be also essential in BEC of atomic gas with large magnetic dipole moments
\cite{Goral2000,Martikainen2001}. Magnetic interactions are usually dominated by van der Waals
forces but effects of magnetic interactions can essentially amplified by reducing of scattering
length $a$ via a Feshbach resonance \cite{Donley2001,Yi2002}. Analysis of this Letter can be
applied for both cases of electric and magnetic dipole-dipole interactions.

In this letter sufficient analytical criteria are developed both for catastrophic collapse of BEC
of a trapped gas of dipolar particles and for long-time condensate existence. Sufficient criteria
allows to predict condensate collapse or, opposite, its long-time existence for given condensate
energy, E, number of particles, N, initial mean square width of condensate, and initial kinetic
energy of condensate. Analytical criteria are compared with results of variational approach
\cite{santosPRL2000}, where collapse was predicted based on the absence of local minimum of ground
state of energy functional provided number of condensate particle exceeds certain  critical value.
 It is shown here that variational calculation gives threshold number
 of particles and condensate energy which are located between parameters
 regions where analytical criteria predict collapse and long-time
 condensate existence, respectively.  It is proven in
this Letter that collapse certainly occurs provided energy of the
condensate exceeds a threshold value which is determined by the
number of particles and trap parameters. Collapse of condensate is
accompanied with dramatic contraction of the atomic cloud.
Collapse is impossible provided number of particles and initial kinetic energy of condensate are
below the critical values.

The time-dependent Gross-Pitaevskii equation (GPE) for atoms with long-range interactions and for
a cylindrical harmonic trap is given by \cite{Yi2000}:
\begin{eqnarray}
\label{GP1}
          i\hbar\frac{\partial\Psi}{\partial t}=          \Big\{ -\frac{\hbar^2\nabla^2}{2m}
           +\frac{1}{2} m\omega_0^2(x_1^2+x_2^2+ \gamma^2x_3^2)
\nonumber\\
+g|\Psi|^2 +\int V({\bf r}-{\bf r}')|\Psi({\bf r}')|^2 d^3{\bf r}'\Big\}\Psi,
\end{eqnarray}
where ${\bf r}=(x_1,x_2,x_3),$ $\Psi$ is the condensate wave
function, coupling constant $g$ corresponds to short-range forces
and  is given by $g=4\pi\hbar^2a/ m$, $a$ is the s-wave
scattering length, $m$ is the atomic mass, $\omega_0$ is a trap
frequency in $x_1x_2$ plane, and $\gamma$ is the anisotropy
factor of the trap. $\Psi({\bf r},t)$ is normalized to the total
number of atoms in condensate: $N=\int |\Psi|^2d^3{\bf r}.$
 It is assumed that  the system is away from shape resonances of $V({\bf
r})$ \cite{Yi2000} and that the long-range potential is due to the
dipole-dipole interaction, and is given by
\begin{equation}
\label{dd1} V({\bf r}-{\bf r}')=
        \frac{[{\bf d}_1({\bf r})\cdot{\bf d}_2({\bf r}')]-3\,
        [{\bf d}_1({\bf r})\cdot{\bf u}]\,[{\bf d}_2({\bf r}')\cdot{\bf u}]}{|{\bf r}- {\bf r}'|^2},
\end{equation}
where ${\bf u}=({\bf r}-{\bf r}')/|{\bf r}-{\bf r}'|$.  All the
dipoles are assumed to point in the direction of trap axes ($\hat
x_3$-direction), i.e., ${\bf d}_1={\bf d}_2=d\hat{x_3}$.
Potential $(\ref{dd1})$ with no dependence of $d$ on ${\bf r}$ is
a good approximation provided a typical interparticle distance
exceeds a few Bohr radii.

GPE $(\ref{GP1})$ can be also obtained from variation of energy
functional, $E:$ $  i\hbar\frac{\partial\Psi}{\partial t}=
\frac{\delta E}{\delta \Psi^*}$, where the condensate energy,
\begin{eqnarray}\label{Edef}
E=E_{K}+E_P+E_{NL}+E_{DD},
\end{eqnarray}
is an integral of motion: $\frac{ d \,E}{d \,t}=0$, and
\begin{eqnarray}\label{Etotdef}
 E_K= \int\frac{\hbar^2}{2m} |\nabla \Psi|^2 d^3{\bf r},
\nonumber\\
E_P=\int \frac{1}{2} m\omega_0^2(x_1^2+x_2^2+
\gamma^2x_3^2)|\Psi|^2d^3{\bf r},
\nonumber\\
E_{NL}=\frac{g}{2}\int |\Psi|^4 d^3{\bf r},
\nonumber\\
E_{DD}=\frac{1}{2}\int |\Psi({\bf r})|^2 V({\bf r}-{\bf
r}')|\Psi({\bf r}')|^2 d^3{\bf r}d^3{\bf r}'.
\end{eqnarray}

Consider time evolution of the mean square radius of the wave
function, $\langle r^2 \rangle \equiv \int r^2 |\Psi|^2 d^3{\bf
r}/N.$ Using $(\ref{GP1}),$ integrating by parts, and taking into
account vanishing boundary conditions at infinity one gets for
the first time derivative
\begin{eqnarray}\label{At}
\partial_t\langle r^2 \rangle=\frac{\hbar}{2mN} \int 2 i x_j  (\Psi\partial_{x_j}\Psi^\ast-\Psi^\ast\partial_{x_j}\Psi)d^3{\bf r},
\end{eqnarray}
where $\partial_{t}\equiv \frac{\partial}{\partial t}$,
$\partial_{x_j}\equiv \frac{\partial}{\partial x_j}$ and repeated
index $j$ means summation over all space coordinates, $j=1,
\ldots, 3$.

In a similar way, after a second differentiation over $t$, one
gets
\begin{eqnarray}\label{Att}
\partial^2_t\langle r^2 \rangle=\frac{1}{2mN}\Big [
8E_K-8E_P+12E_{NL} \nonumber\\
-2\int|\Psi({\bf r}|^2 |\Psi({\bf r'}|^2(x_j
\partial_{x_j}+x'_j\partial_{x'_j})V({\bf r}-{\bf r}')d^3{\bf r}\Big ].
\end{eqnarray}
Note that, in the case $E_P=0, \ V({\bf r})\equiv 0$, Eq.
$(\ref{Att})$ coincides with the so-called virial theorem for the
GPE with local interactions
\cite{vlas,zakh1972,lush1995,Pitaevskii1995,LushnikovSaffman2000}
thus it is natural to call Eq. $(\ref{Att})$ by a virial theorem
for GPE $(\ref{GP1})$.

Using Eq. $(\ref{dd1})$ one gets $(x_j
\partial_{x_j}+x'_j\partial_{x'_j})V({\bf r}-{\bf r}')=-3V({\bf r}-{\bf
r}')$ and using  Eq. $(\ref{Edef})$ one can rewrite virial theorem $(\ref{Att})$ as follows:
\begin{eqnarray}\label{Att2}
\partial^2_t\langle r^2 \rangle=\frac{1}{2mN}\Big
[12E-4E_K-10m\omega_0^2N\langle r^2\rangle
\nonumber\\
-10m\omega_0^2N(\gamma^2-1)\langle x_3^2\rangle\Big ].
\end{eqnarray}
It is essential here that both local nonlinear term and nonlocal term are included into the energy
$E$ which is a conserved quantity. Catastrophic collapse of BEC occurs while $\langle
r^2\rangle\to 0$. From mathematical point of view it means that if, according to virial theorem
$(\ref{Att2})$, the positive-definite quantity $\langle r^2\rangle$ becomes negative in finite
time then singularity in solution of Eq. $(\ref{GP1})$ appears in a finite time before $\langle
r^2\rangle$ becomes negative and singularity in solution of GPE occurs together with catastrophic
squeezing of the distribution of $|\Psi|.$ Near singularity formation GPE is not applicable and
another physical mechanisms are important such as inelastic two- and three-body collisions which
can cause a loss of atoms from the condensate \cite{PitaevskiiRevModPhys1999}. In addition, long
term interactions are described by the dipole-dipole potential $(\ref{dd1})$ provided typical
distance between atoms in condensate exceeds a few Bohr radii. Note that regularization of
potential $(\ref{dd1})$ to avoid singularity at ${\bf r}=0$ allows to prevent singularity
formation in GPE \cite{Turitsyn1985,Perez2000}.
 However GPE $(\ref{GP1})$ can still
describe significant contraction of atomic cloud.

Thus condition $\langle r^2\rangle \to 0$ provides a sufficient
criterion of collapse of BEC. E.g. one immediately obtains from
Eq. $(\ref{Att2})$ that $\partial^2_t\langle r^2
\rangle<\frac{6E}{mN}$ and collapse is inevitable for $E<0$. One
can obtain however much more strict sufficient condition for
collapse using generalized uncertainty relations between $E_K, \
N, \langle r^2\rangle, \ \partial _t\langle r^2\rangle$
\cite{lush1995} which follows from Cauchy-Schwarz inequality and
Eq. $(\ref{At})$  with use of integration by parts $(\Psi\equiv
Re^{i\phi},$ $R=|\Psi|)$:
\begin{eqnarray}\label{heis1}
E_K=\frac{\hbar^2}{2m}\int\Big [(\nabla R)^2+(\nabla \phi)^2R^2\Big ]d^3{\bf r},
\nonumber \\
\frac{2mN}{\hbar}|\partial _t\langle r^2\rangle|=4|\int x_j\partial_{x_j}\phi R^2d^3{\bf r}| \nonumber \\
\le
4\Big (N\langle r^2\rangle\int(\nabla \phi)^2R^2d^3{\bf r}\Big)^{1/2},
\nonumber \\
N=-\frac{2}{3}\int{x_j}R\partial _{x_j}Rd^3{\bf r}\le \frac{2}{3}\Big (N\langle
r^2\rangle\int(\nabla R)^2d^3{\bf r}\Big)^{1/2}.
\end{eqnarray}
Using Eqs. $(\ref{Att2}),(\ref{heis1})$ one can obtain a basic differential inequality:
\begin{eqnarray}\label{ineq1}
\partial^2_t\langle r^2 \rangle\le
\frac{1}{2mN}\Big [12E-\frac{\hbar^2}{2m}\Big( \frac{9N}{\langle r^2
\rangle}+\frac{m^2N(\partial_t\langle r^2 \rangle)^2}{\hbar^2\langle r^2 \rangle}\Big )
 \nonumber \\
-10m\omega_0^2NF(\gamma)\langle r^2\rangle\Big ],
\end{eqnarray}
where $F(\gamma)=1$ for $\gamma\ge 1$ and $F(\gamma)=\gamma^2$
for $\gamma<1$. Change of variable, $\langle r^2
\rangle=B^{4/5}/N$ gives the differential inequality:
\begin{eqnarray}\label{Bineq1}
\partial^2_t B\le
\frac{5}{2m}\Big
[3EB^{1/5}-\frac{\hbar^2}{8m}\frac{9N^2}{B^{3/5}}-
\frac{5}{2}m\omega_0^2F(\gamma)B\Big ],
\end{eqnarray}
which can be rewritten as
\begin{eqnarray}\label{Btt12}
     B_{tt} = -\frac{\partial U(B)}{\partial B}-f^2(t),
\end{eqnarray}
%
where
\begin{eqnarray}\label{Udef}
    U=
    -\frac{25}{4m}EB^{6/5}+\frac{\hbar^2225N^2}{32m^2}B^{2/5}+
\frac{25}{8}\omega_0^2F(\gamma)B^2,
\end{eqnarray}
\noindent and $f^2(t)$ is some unknown nonnegative function of
time. Equation $(\ref{Btt12})$ has a simple mechanical analogy
\cite{lush1995}  with the motion of a ``particle" with coordinate
$B$ under the influence of the  potential force $-\frac{\partial
U(B)}{\partial B}$ in addition to the force $-f^2(t)$. Due to the
influence of the nonpotential force $-f^2(t)$ the total energy
$\mathcal E$ of the ``particle" is time dependent: ${\mathcal
E}(t)=\frac{B_t^2}{2}+U(B)$. Collapse certainly occurs if the
``particle" reaches the origin $B=0.$ It is clear that if the
particle were to reach the origin without the influence of the
force $-f^2(t)$ then it would reach the origin even faster under
the additional influence of this nonpositive force. Therefore one
can consider below the particle dynamics  without the influence
of  the nonconservative force $-f^2(t)$ to prove sufficient
collapse conditions.

It follows from Eq. $(\ref{Udef})$ that potential $U(B)$ is a
monotonic function for $E\le \hbar \omega_0
N[F(\gamma)5]^{1/2}/2\equiv E_{critical}$ (see curve 1 in Fig. 1)
\begin{figure}
\begin{center}
\includegraphics[width = 3.4 in]{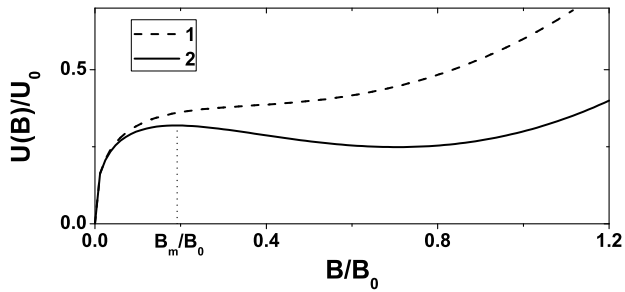}
\caption{Typical behaviour of potential $U(B)$ from Eq. $(\ref{Udef})$ for $E\le E_{critical}$
(curve 1)  $E> E_{critical}$ (curve 2). $U_0=(N^5\hbar^5/m^5 \omega_0)^{1/2}, \quad
B_0=(N\hbar/m\omega_0)^{5/4}.$} \label{fig:fig1}
\end{center}
\end{figure}
while for $E>E_{critical}$ potential $U(B)$ has a barrier at
$B_m^{4/5}=3\Big (E-\big [E^2-E_{critical}^2\big
]^{1/2}\Big)/\big[5 m\omega_0^2F(\gamma)\big]$ with particle
energy ${\mathcal E}_m=U(B_m)$ at the top (see curve 2 in Fig.
1). One can separate sufficient collapse condition into three
different cases:

(a)  for $E\le E_{critical}$
 the particle reaches the origin in a finite time irrespective of the initial value of $B\bigl|_{t=0}$;

(b) for $E> E_{critical}$ and ${\mathcal E}(0)>{\mathcal E}_m$, the particle is able to overcome
the barrier thus it always falls to the origin in a finite time irrespective of the initial value
of $B\bigl|_{t=0}$;

(c) for $E> E_{critical}$ and ${\mathcal E}(0)<{\mathcal E}_m$, the particle is not able to
overcome the barrier thus it falls to the origin in a finite time only if $B\bigl|_{t=0}<B_m.$

Note that it is proven here analytically only sufficient collapse conditions. It means that even
if none of conditions a,b,c are satisfied one can not exclude collapse formation for some
particular values of the initial conditions of Eq. $(\ref{GP1}).$ Generally it is determined by
nonpotential force  $-f^2(t)$. However inequality $(\ref{ineq1})$ reduces to equality for a
Gaussian initial condition and $\gamma=1$:
\begin{eqnarray}\label{psi0}
   \Psi_0=N^{1/2}\pi^{-3/4}(L_\rho^2 L_3)^{-1/2}e^{-(x_1^2+x_2^2)/2L_\rho^2}e^{-x_3^2/2L_3^2}
\end{eqnarray}
in particular case with $L_3=L_\rho$.

One can compare sufficient collapse condition with results of
Ref. \cite{santosPRL2000} where collapse was predicted from
variational analysis using Gaussian ansatz $(\ref{psi0})$ to
approximate ground state of GPE for $g=0$ $(\ref{GP1})$. It was
concluded that collapse should occur provided energy functional
$E$ has no local minima. A critical point was determined from the
condition that local minimum of energy functional $E$ becomes a
saddle point: $(\partial^2 E/\partial L_{3}^2)(\partial^2
E/\partial L_{\rho}^2)=(\partial^2 E/\partial L_{3}\partial
L_\rho)^2, \
\partial E/\partial L_{3}=\partial^2 E/\partial L_{\rho}=0.$ This
allows to find critical number of particle, $N_{c,\, var}$ and
critical value of energy functional, $E_{c,\, var}$, as a function
of system parameters $L_\rho, \, L_3, \, \gamma, \, d.$. Fig. 2
shows dependence of $E_{c,\, var}$ (curve 1) and $E_{critical}$
(curve 2) on trap aspect ratio $\gamma$ for $N=N_{c,\, var}$ and
$g=0$.
\begin{figure}
\begin{center}
\includegraphics[width = 3.4 in]{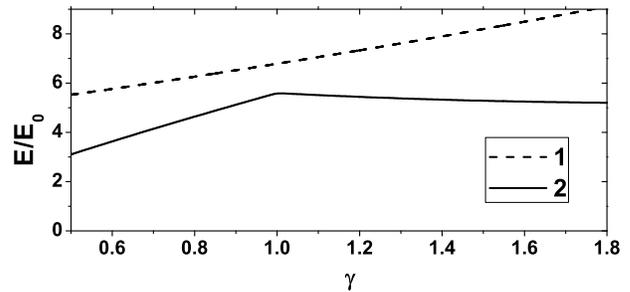}
\caption{Dependence of $E_{c,\, var}$ (curve 1) and $E_{critical}$ (curve 2) on trap aspect ratio
$\gamma$ for $N=N_{c,\, var}$. Both $E_{c,\, var}$ and $E_{critical}$ are given in units of
$E_0=\hbar^{7/2}\omega_0^{1/2}/d^2m^{3/2}$. } \label{fig:fig2}
\end{center}
\end{figure}
Note that the expression for $E_{c,\, var}$, used in this Letter to draw curve 2 in Fig. 2, differs
from Eq. $(3)$ of Ref. \cite{santosPRL2000}. The authors of Ref. \cite{santosPRL2000} already
mentioned in the erratum \cite{santosPRL2002E}) that Eq. $(3)$ of \cite{santosPRL2000} is
incorrect. However a corrected formula was not given in the erratum \cite{santosPRL2002E}.
Explicit expression for $E_{c,\, var}$ is not given in this Letter also because it
 is very bulky and will be given elsewhere.
 Gaussian ansatz $(\ref{psi0})$ is
not an exact solution of GPE $(\ref{GP1})$ thus one can expect that actual critical value of
energy $E$ of ground state solution is lower. $E_{critical}$ is determined here from sufficient
collapse condition meaning that critical value of energy $E$ of ground state solution is always
above curve 2. One can conclude that actual critical value of energy is located between curves 1
and 2. Accuracy of variation approximation can generally be obtained only from comparison with
direct simulation of GPE $(\ref{GP1})$.


 The Fourier transform of dipole-dipole
interaction potential  $(\ref{dd1})$ allows to find a sufficient
condition of global existence (for arbitrary large time) of
solution of GPE $(\ref{GP1})$. The dipole-dipole interaction
energy $E_{DD}$ can be rewritten in ${\bf k}$-space as
$E_{DD}=(1/2)\int |R_{\bf k}|^2 V_{\bf k} d^3{\bf k}/(2\pi)^3$,
where $R_{\bf k}$ is a Fourier transform of $|\Psi|^2$ and the
Fourier transform of the dipole-dipole interaction in the limit
of small atomic overlap distance is given by~\cite{Goral2000}: $
V_{\bf k}=-\frac{4\pi}{3}d^2(1-3\cos^2\alpha).$
Here $\alpha$ is the angle between ${\bf k}$ and ${\bf d}$. Using inequality $ V_{\bf k}\ge
-\frac{4\pi}{3}d^2$ one gets $E_{DD}\le-\frac{2\pi}{3}d^2 Y$, where $Y\equiv \int |\Psi|^4 d^3
{\bf r}$. Condition $4\pi d^2\le 3g$ results in $E>0$ for any particle number and collapse is
impossible. Below it is assumed that $4\pi d^2>3g$. $Y$ can be bounded as follows: $Y\le
\frac{4}{3^{3/2}N_0}N^{1/2}X^{3/2},$ where $X\equiv \int |\nabla \Psi|^2 d^3 {\bf r}$, and
$N_0=18.94$ is determined from a ground state solution, $\phi_0=\lambda R(\lambda {\bf r}) e^{i
\lambda^2 t}$, of nonlinear Schr\"odinger equation: $-\lambda^2 R+\nabla^2 R+R^3=0,$ $N_0\equiv
\int R^2d^3 {\bf r}$ (see Ref. \cite{KuznetsovRasmussen1995}).  Using these inequalities and Eqs.
$(\ref{Edef}),(\ref{Etotdef}),(\ref{heis1})$ one gets lower bound of energy functional:
\begin{eqnarray}\label{Eineq}
   E\ge\frac{\hbar^2}{2m}X+
   \frac{9m\omega^2}{8X}F(\gamma)N^2-\frac{2(4\pi d^2-3g)}{3^{5/2}N_0}
   N^{1/2}X^{3/2}\nonumber \\
   \equiv E_l(X).
\end{eqnarray}

For $N>N_c,$ $N_c\equiv 2^{3/2}3\hbar^{5/2}N_0/[5^{5/4} (4\pi
d^2-3g) F(\gamma)^{1/4}  m^{3/2}\omega^{1/2}],$ the function
$E_l(X)$ is a monotonic one (curve 1 in Fig. 3). For $N<N_c$ the
function $E_l(X)$ has a local minimum, $E_{min}$ (curve 2 in Fig.
3).
\begin{figure}
\begin{center}
\includegraphics[width = 3.8 in]{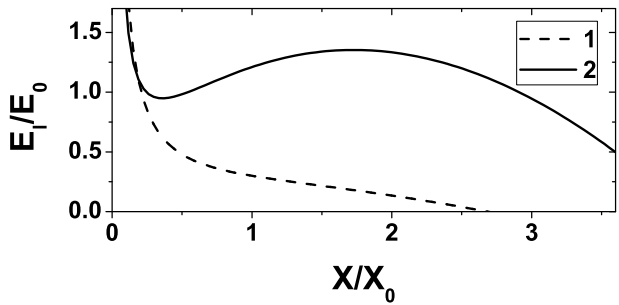}
\caption{Energy lower bound $E_l$ (in units of $E_0$) for $N>N_c$
(curve 1) and $N<N_c$ (curve 2) versus $X/X_0$. $X_0=N^{3/5}\big [
mN_0\omega_0^2/(4\pi d^2-3g)\big ]^{2/5}$.} \label{fig:fig3}
\end{center}
\end{figure}
Consider initial condition with
\begin{equation} \label{X1X2}
N<N_c, \ E_{min}<E<E_{max}, \ X_1<X|_{t=0}<X_2,
\end{equation}
where $E_{max}$ is a local maximum of $E_l(X)$, and $X_1, X_2$ are two of total three roots
$(X_1<X_2<X_3)$ of Eq. $E=E_l(X)$. Any solution of GPE, corresponding to conditions $(\ref{X1X2})$,
will stay in the range $X_1<X<X_2$ at any time because regions below curves 1,2 in Fig. 3 are
forbidden for solution of GPE. One concludes that collapse is impossible in that case because
collapse and singularity formation in GPE requires singularity in kinetic energy
\cite{Weinstein1983}, $X\to \infty$. That could be understood e.g. from uncertainty relations
$(\ref{heis1})$. Eq. $(\ref{X1X2})$ gives a sufficient condition of absence of collapse and in
that case one can expect that energy functional $E$ has a local minimum and supports stable
steady-state solutions.
 In
original quantum mechanical problem that steady state is
metastable one because of finite probability of tunneling of
condensate from local minimum which is outside the applicability
of GPE and is not considered in this Letter.

In conclusion, sufficient analytical criteria are developed both
for catastrophic collapse of BEC of gas with nonlocal long-range
dipole-dipole interactions and for long-time collapse existence
in the framework of GPE $(\ref{GP1})$.


The author thanks M. Chertkov and I.R. Gabitov for helpful discussions.

Support was provided by the Department of Energy, under contract W-7405-ENG-36.




\end{document}